\documentclass[prd,showpacs,eqsecnum,floatfix]{revtex4}%
\usepackage{amsmath}
\usepackage{graphicx}
\usepackage{pstricks}%
\usepackage{amsfonts}%
\usepackage{amssymb}

\newcommand{\BOX}{\hbox {$\sqcap$ \kern -1em $\sqcup$}}

\newcommand{\be}{\begin{equation}}
\newcommand{\ee}{\end{equation}}
\newcommand{\ba}{\begin{eqnarray}}
\newcommand{\ea}{\end{eqnarray}}
\newcommand{\ban}{\begin{eqnarray*}}
\newcommand{\bea}{\begin{eqnarray}}
\newcommand{\eea}{\end{eqnarray}}
\newcommand{\ean}{\end{eqnarray*}}
\newcommand{\barr}{\begin{array}}
\newcommand{\earr}{\end{array}}

\begin{document}

\title{Decoherence induced CPT\ violation and
entangled neutral mesons}
\author{J.~Bernab\'eu}
\affiliation{Departamento de F\'isica Te\'orica and IFIC, Universidad de
Valencia-CSIC, E-46100, Burjassot (Valencia), Spain.}
\author{Nick~E.~Mavromatos and Sarben~Sarkar}
\affiliation{King's College London, University of London, Department of Physics, Strand
WC2R 2LS, London, U.K.}

\begin{abstract}
{\small We discuss two classes of semi-microscopic theoretical models of
stochastic space-time foam in quantum gravity and the associated effects on
entangled states of neutral mesons, signalling an intrinsic breakdown of CPT
invariance. One class of models deals with a specific model of foam, initially
constructed in the context of non-critical (Liouville) string theory, but
viewed here in the more general context of effective quantum-gravity models.
The relevant Hamiltonian perturbation, describing the interaction of the meson
with the foam medium, consists of off-diagonal stochastic metric fluctuations,
connecting distinct mass eigenstates (or the appropriate generalisation
thereof in the case of K-mesons), and it is proportional to the relevant
momentum transfer (along the direction of motion of the meson pair). There are
two kinds of CPT-violating effects in this case, which can be experimentally
disentangled: one (termed ``$\omega$-effect'') is associated with the failure
of the indistinguishability between the neutral meson and its antiparticle,
and affects certain symmetry properties of the initial state of the two-meson
system; the second effect is generated by the time evolution of the system in
the medium of the space-time foam, and can result in time-dependent
contributions of the $\omega$-effect type in the time profile of the two meson
state. Estimates of both effects are given, which show that, at least in
certain models, such effects are not far from the sensitivity of  
experimental facilities available currently or in the near future. The other class of quantum gravity
models involves a medium of gravitational fluctuations which behaves like a
``thermal bath''. In this model both of the above-mentioned intrinsic CPT
violation effects are not valid.}

\end{abstract}
\maketitle


\section{Introduction and Motivation}

There have been two strands of research in the last few years which have only
recently made contact. One is the subject of bipartite entanglement and the
other is the role of space-time foam \ for decoherence of elementary
particles. The latter was first championed by Wheeler within the context of
microscopic horizons of radius of the order of the Planck length which may
induce in space-time a fuzzy structure. This has been further developed in
\cite{hawking} where it was suggested that topological fluctuations in the
space-time background \ and microscopic black holes can lead to non-unitary
evolution and a breakdown of the S-matrix description in field theory. If this
is correct then the usual formulation of quantum mechanics has to be modified.
Arguments have been put forward for this modification of the Liouville
equation to take the form~\cite{ehns}
\begin{equation}
\partial_{t}\rho=\frac{i}{\hbar}\left[  \rho,H\right]  +\not \delta H\rho.\,
\label{liouvqm}%
\end{equation}
Equations of this form are frequently required to describe the time evolution
of an open quantum mechanical system where $\not \delta H\rho$ has a Lindblad
form~\cite{lindblad}. In such systems observable degrees of freedom are
coupled to unobservable components which are effectively integrated over.
Initial pure states evolve into mixed ones and so the S-matrix $\not S  $
relating initial and final density matrices does not factorise, i.e.
\begin{equation}
\not S  \neq SS^{\dag}%
\end{equation}
where $S=e^{iHt}$. In these circumstances Wald~\cite{wald} has shown that CPT
is violated, at least in its strong form, i.e. there is \textit{no} unitary
invertible operator $\Theta$ such that
\begin{equation}
\Theta\overline{\rho}_{\mathrm{in}}=\rho_{\mathrm{out}}\mathrm{.}%
\end{equation}
This result is due to the entanglement of the gravitational fluctuations with
the matter system. Such entanglement is not generally a perturbative effect.

It was pointed out in \cite{Bernabeu}, that if the CPT operator is not well
defined this has implications for the symmetry structure of the initial
entangled state of two neutral mesons in meson factories. Indeed, if CPT can
be defined as a quantum mechanical operator, then the decay of a (generic)
meson with quantum numbers $J^{PC} = 1^{ - - } $ \cite{Lipkin}, leads to
a pair state of neutral mesons $\left|  i\right\rangle $ having the form of
the entangled state%
\begin{equation}
\left|  i\right\rangle =\frac{1}{\sqrt{2}}\left(  \left|  \overline{M_{0}%
}\left(  \overrightarrow{k}\right)  \right\rangle \left|  M_{0}\left(
-\overrightarrow{k}\right)  \right\rangle -\left|  M_{0}\left(
\overrightarrow{k}\right)  \right\rangle \left|  \overline{M_{0}}\left(
-\overrightarrow{k}\right)  \right\rangle \right)  .
\end{equation}
This state has the Bose symmetry associated with 
particle-antiparticle indistinguishability
$C{\cal P}=+$, where $C$ is the charge conjugation and ${\cal P}$ is 
the permutation operation. 
If, however, CPT is not a good symmetry (i.e. ill-defined due
to space-time foam), then $M_{0}$ and $\overline{M_{0}}$ may not be identified
and the requirement of $C{\cal P}=+$ is relaxed~\cite{Bernabeu}. Consequently,
in a perturbative 
framework,
the
state of the meson pair can be parametrised to have the following form: 
\begin{align*}
\left|  i\right\rangle  &  =\frac{1}{\sqrt{2}}\left(  \left|  \overline{M_{0}%
}\left(  \overrightarrow{k}\right)  \right\rangle \left|  M_{0}\left(
-\overrightarrow{k}\right)  \right\rangle -\left|  M_{0}\left(
\overrightarrow{k}\right)  \right\rangle \left|  \overline{M_{0}}\left(
-\overrightarrow{k}\right)  \right\rangle \right) \\
&  +\frac{\omega}{\sqrt{2}}\left(  \left|  \overline{M_{0}}\left(
\overrightarrow{k}\right)  \right\rangle \left|  M_{0}\left(  -\overrightarrow
{k}\right)  \right\rangle +\left|  M_{0}\left(  \overrightarrow{k}\right)
\right\rangle \left|  \overline{M_{0}}\left(  -\overrightarrow{k}\right)
\right\rangle \right)
\end{align*}
where $\omega=\left|  \omega\right|  e^{i\Omega}$ is a complex CPT violating
(CPTV) parameter. For definiteness in what follows we shall term this
quantum-gravity effect in the initial state as the ``$\omega$%
-effect''\cite{Bernabeu}.

Given the possible breakdown of conventional unitary quantum mechanics and
non-invariance of CPT as a consequence of space-time foam, it is interesting
to see what, if any, type of foam can generate this type of effect from a
fundamental and non-phenomenological stance. This is a somewhat delicate
matter since evolution using the popular Lindblad approaches 
\cite{lindblad},\cite{kaons},\cite{waldron} 
generate effects, which, however,
have essential differences in both form and interpretation from the $\omega
$-effect'', and thus 
can be experimentally disentangled from it~\cite{waldron}. 
However the Lindblad-type approach to quantum gravity is primarily based on
mathematical considerations of quantum dynamical semigroups (required for
irreversibility) and Markov processes and does not claim any other physical
motivation. It is certainly not a microscopic theory of quantum gravity. There
is need for a more detailed microscopic and model dependent approach in order
to arrive at reasonable estimates for the $\omega$-(initial state) and its
evolution. In this work we shall study and estimate such effects in the
context of specific models of space-time foam which are motivated by
underlying theoretical considerations.

Space-time foam is a generic term that covers quite distinct points of view.
One of the most plausible reasons for considering quantum decoherence models
of quantum gravity, comes from recent astrophysical evidence for the
acceleration of our Universe during the current-era. Observations of distant
supernovae~\cite{snIa}, as well as WMAP data~\cite{wmap} on thermal
fluctuations in the cosmic microwave background (CMB), indicate that our
Universe is at present accelerating, and that 74\% of its energy-density
budget consists of an unknown entity, termed \emph{Dark Energy}. Best-fit
models of such data include Einstein-Friedman-Robertson-Walker Universes with
a non-zero cosmological \emph{constant}. However, the data are currently
compatible also with (cosmic) time-dependent vacuum-energy-density components,
relaxing asymptotically to zero~\cite{steinhardt}. Such relaxation mechanisms
may be due to extra scalar \emph{quintessence} fields~\cite{quintess}, which
in the case of some models inspired by non-critical string theory might be the
dilaton itself~\cite{emngrf}. Identification of the dark energy component of
the Universe with the central charge surplus of the supercritical $\sigma
$-models describing the (recoil) string excitations of colliding brane-worlds
leads to a non-equilibrium energy density of the (observable) brane
world~\cite{gravanis,emw}and a relaxation scenario for the dark energy. The
associated dilaton field during the present era of the Universe may even be
constant, in which case the relaxation of the dark-energy density component is
a purely stringy feature of the logarithmic conformal field
theory~\cite{lcftmav} describing the D-brane recoil~\cite{kmw} in a
(perturbative) $\sigma$-model framework. This is compatible with a de Sitter
space, with scale factor $a(t)=\mathrm{exp}(\sqrt{\Omega_{0}/3}t)$, which
implies an asymptotic Hubble horizon
\begin{equation}
\delta_{H}\sim\int_{t_{0}}^{\infty}d(ct^{\prime})a^{-1}(t^{\prime}%
)~<~\infty\label{horizon}%
\end{equation}

One suggestion for the quantisation of such systems is through analogies with
open systems in quantum mechanics. For some simple cases, such as conformally
coupled scalar fields~\cite{coarse} in de Sitter spaces it has been shown
explicitly that the system modes decohere for wavelengths longer than a
critical value, which is of the order of the Hubble horizon. From the theorem
by Wald~\cite{wald}, the CPT operator is ill-defined in such decoherent field
theories. Non-critical (Liouville) string~\cite{ddk} provides a rather unified
formalism for dealing not only with cosmological constant Universes in string
theory, but also in general with decoherent quantum space-time foam
backgrounds, that include microscopic quantum-fluctuating black
holes~\cite{emn}. Within this framework a particularly simple and tractable
background is given by D-particles. Low energy matter is represented as open
or closed strings \ and moves in a $D+1$ dimensional target space. The string
states collide with massive D-particle defects embedded in target space. The
recoil fluctuations of the D-particle induce a space-time distortion given by
the metric tensor
\begin{equation}
g_{ij}=\delta_{ij},\,g_{00}=-1,g_{0i}=\varepsilon\left(  \varepsilon
y_{i}+u_{i}t\right)  \Theta_{\varepsilon}\left(  t\right)  ,\;i=1,\ldots,D
\label{recmetr}%
\end{equation}
where the suffix $0$ denotes temporal (Liouville) components and
\begin{align}
\Theta_{\varepsilon}\left(  t\right)   &  =\frac{1}{2\pi i}\int_{-\infty
}^{\infty}\frac{dq}{q-i\varepsilon}e^{iqt},\label{heaviside}\\
u_{i}  &  =\left(  k_{1}-k_{2}\right)  _{i}\;,\nonumber
\end{align}
with $k_{1}\left(  k_{2}\right)  $ the momentum of the propagating
closed-string state before (after) the recoil; $y_{i}$ are the spatial
collective coordinates of the D particle and $\varepsilon^{-2}$ is identified
with the target Minkowski time $t$ for $t\gg0$ after the collision~\cite{kmw}.
These relations have been calculated for non-relativistic branes where $u_{i}$
is small. Now for large $t,$ to leading order,%
\begin{equation}
g_{0i}\simeq\overline{u}_{i}\equiv\frac{u_{i}}{\varepsilon}\propto\frac{\Delta
p_{i}}{M_{P}} \label{recoil}%
\end{equation}
where $\Delta p_{i}$ is the momentum transfer during a collision and $M_{P}$
is the Planck mass (actually, to be more precise $M_{P}=M_{s}/g_{s}$, where
$g_{s}<1$ is the (weak) string coupling, and $M_{s}$ is a string mass scale);
so $g_{0i}$ is constant in space-time but depends on the energy content of the
low energy particle \cite{Dparticle}. Such a feature does not arise in
conventional approaches to space-time foam and will be important in our
formulation of one of the microscopic models that we will consider.

The above model of space-time foam refers to a specific string-inspired
construction. However the form of the induced back reaction (\ref{recoil})
onto the space-time has some generic features, and can be understood more
generally in the context of effective theories of such models, which allows
one to go beyond a specific non-critical (Liouville) model. Indeed, the
D-particle defect can be viewed as an idealisation of some (virtual, quantum)
black hole defect of the ground state of quantum gravity, viewed as a membrane
wrapped around some small extra dimensions of the (stringy) space time, and
thus appearing to a four-dimensional observer as an ``effectively'' point like
defect. The back reaction on space-time due to the interaction of a pair of
neutral mesons, such as those produced in a meson factory, with such defects
can be studied generically as follows: consider the non-relativistic recoil
motion of the heavy defect, whose coordinates in space-time,in the laboratory
frame, are $y^{i} = y^{i}_{0} + u^{i} t$, with $u^{i} $ the (small) recoil
velocity. One can then perform a (infinitesimal) general coordinate
transformation $y^{\mu}\to x^{\mu}+ \xi^{\nu}$ so as to go to the rest (or
co-moving ) frame of the defect after the scattering. From a passive point of
view, for one of the mesons, this corresponds to an induced change in metric
of space-time of the form (in the usual notation, where the parenthesis in
indices denote symmetrisation) $\delta g_{\mu\nu} = \partial_{(\mu}\xi_{\nu)}%
$, which in the specific case of non-relativistic defect motion yields the
off-diagonal metric elements (\ref{recoil}).
Such transformations cannot be performed simultaneously for both mesons, 
and moreover in a full theory of quantum gravity the recoil 
velocities fluctuate randomly, as we shall discuss later on.
This means that the effects 
of the recoil of the
space-time defect are observable. 
The mesons will feel such effects 
in the form of induced fluctuating metrics (\ref{recoil}). It
is crucial to note that the interaction of the matter particle (meson) with
the foam defect may also result in a ``flavour''
change of the particle (e.g. the change of a neutral meson to its
antiparticle). This feature can be understood in a D-particle Liouville model
by noting that the scattering of the matter probe off the defect involves
first a splitting of a closed string representing matter into two open ones,
but with their ends attached to the D-particle, and then a joining of the
string ends in order to re-emit a closed string matter state. The re-emitted
(scattered) state may in general be characterised by phase, flavour and other
quantum charges which may not be required to be conserved during black hole
evaporation and disparate space-time-foam processes. In our application we
shall restrict ourselves only to effects that lead to flavour changes. The
modified form of the metric fluctuations (\ref{recoil}), which characterise
our specific model of Liouville decoherence is given in the next section.

The second model to be considered by us shares a concern with the effect of
horizons and the consequent absence of unitarity but the formulation is not
supported by a formal theory like string theory. A different effective theory
of space-time foam has been proposed by Garay \cite{Garay}. The fuzziness of
space-time \ at the Planck scale is described by a non-fluctuating background
which is supplemented by non-local interactions. The latter reflects the fact
that at Planck scales space-time points lose their meaning and so these
fluctuations present themselves in the non-fluctuating coarse grained
background as non-local interactions. These non-local interactions are then
rephrased as a quantum thermal bath with a Planckian temperature. The quantum
entanglement of the gravitational bath and the two meson (entangled) state is
explicit in this model. Consequently issues of back reaction can be readily
examined. Since the evolution resulting from the standard Lindblad formulation
does not lead to the $\omega$ effect, this manifestation of CPTV is not the
result of some arbitrary non-unitary evolution. Hence it is interesting to
study the above two quite distinct models (one motivated by string theory and
the other by field theory) for clues concerning the appearance of (and an
estimate for the order of magnitude) of $\omega$.

The structure of this article is as follows: In the next section we will
construct the model incorporating ideas from Liouville string theory that were
mentioned earlier. It will be referred to as the Liouville stochastic metric
(LSM) model. It will be demonstrated that this model leads to the appearance of
$\omega$-type effects both in the initial state and during evolution in the
foam medium. In fact, there are some subtleties 
concerning strangeness conservation (on choosing the $M$ meson to be the $K$ meson for definiteness);
consequently, the precise form 
of the pertinent CPT Violating terms in the 
decay products of a meson factory, constitutes a sensitive probe of 
any non-conservation. In models of space-time foam such as LSM
strangeness is not always conserved, as there is no
corresponding no-hair theorem for the associated (singular) space-time 
fluctuations. This is a feature that is model dependent, and, 
in this work, 
we pay particular attention to determining the conditions under which 
this happens.
Qualitative estimates of the effects are given and the r\^ole of
time dependence in disentangling the effects is discussed. The neutral meson
system and the thermal bath, representing space-time foam, together form one
large hamiltonian system whose evolution can be calculated exactly. This
implies in principle an exact knowledge of the dynamics of the neutral mesons.
Consequently in the third section, entanglement that is induced in the state
of the neutral meson pair by the thermal bath, is calculated. However, as
discussed there, no $\omega$ type effect is generated by the evolution in such
a type of foam. Moreover owing to the details of the entanglement of bath and
neutral mesons the stationary states of the system cannot be interpreted in
terms of an initial $\omega$ effect. Conclusions are presented in the final
section. Technical aspects of our work are given in two appendices.

\bigskip

\section{Liouville inspired decoherence}

\subsection{Liouville-Stochastic-Metric (LSM) fluctuations and Meson systems: formalism}

Polchinski's realisation~\cite{polch} that solitonic string backgrounds
(D-branes) can be described in a conformally invariant way in terms of world
sheets with boundaries has significantly changed the understanding of target
space structure. Collective target space coordinates of the soliton have
Dirichlet boundary conditions on these boundaries. A model of space-time foam
\cite{emw} can be based on a number (determined by target space supersymmetry)
of parallel brane worlds with three large spatial dimensions which move in a
bulk space-time containing a ``gas'' of D-particles. One of these branes is
the observable Universe. For an observer on the brane the crossing D-particles
will appear as twinkling space-time defects, i.e. microscopic space-time
fluctuations. This will give the four-dimensional brane world a ``D-foamy''
structure. Following some recent work on gravitiational decoherence
\cite{NickSarben},\cite{GravDecoh} the target space metric state, which is
close to being flat, can be represented schematically as a density matrix
\begin{equation}
\rho_{\mathrm{grav}}=\int d\,^{5}r\,\,f\left(  r_{\mu}\right)  \left|
g\left(  r_{\mu}\right)  \right\rangle \left\langle g\left(  r_{\mu}\,\right)
\right|  .\, \label{gravdensity}%
\end{equation}
\bigskip The parameters $r_{\mu}\,\left(  \mu=0,\ldots,5\right)  $ \ are
stochastic with a gaussian distribution $\,f\left(  r_{\mu}\,\right)  $
characterised by the averages%
\[
\left\langle r_{\mu}\right\rangle =0,\;\left\langle r_{\mu}r_{\nu
}\right\rangle =\Delta_{\mu}\delta_{\mu\nu}\,.
\]
The fluctuations experienced by the two entangled neutral mesons will be
assumed to be independent and $\Delta_{\mu}\sim O\left(  \frac{E^{2}}%
{M_{P}^{2}}\right)  $i.e. very small. As matter moves through the space-time
foam in a typical ergodic picture the effect of time averaging is assumed to
be equivalent to an ensemble average. As far as our present discussion is
concerned we will consider a semi-classical picture for the metric and so
$\left|  g\left(  r_{\mu}\right)  \right\rangle $ in \ref{gravdensity} will be
a coherent state. In the future we will also address non-classical
fluctuations where the $\left|  g\left(  r_{\mu}\right)  \right\rangle $ could
represent squeezed states of gravitons.

In order to address oscillation and MSW-like
phenomena~\cite{NickSarben}~\cite{GabNick},\cite{MSW} the fluctuations
of each component of the metric tensor $g^{\alpha\beta}$ will not be
simply given by the simple recoil distortion (\ref{recoil}), but
instead will be taken to have a $2\times2$ (``flavour'') structure:
\begin{align}
g^{00}  &  =\left(  -1+r_{4}\right)  \mathsf{1}\nonumber\\
g^{01}  &  =g^{10}=r_{0}\mathsf{1}+ r_{1}\sigma_{1}+ r_{2}\sigma_{2}%
+r_{3}\sigma_{3}\label{metric}\\
g^{11}  &  =\left(  1+r_{5}\right)  \mathsf{1}\nonumber
\end{align}
where $\mathsf{1}$ , is the identity and $\sigma_{i}$ are \ the Pauli
matrices. The above parametrisation has been taken for simplicity and we will
also consider motion to be in the $x$- direction which is natural since the
meson pair moves collinearly in the Center-of-Mass (C.M.) 
frame. A metric with this type of structure is
compatible with the view that the D-particle defect is a ``point-like''
approximation for a compactified higher-dimensional brany black hole, whose no
hair theorems permit non-conservation of flavour. In the case of neutral
mesons the concept of ``flavour''
refers to either particle/antiparticle species or the two mass eigenstates,
by changing appropriately the relevant coefficients.

The Klein-Gordon equation for a spinless neutral meson field $\Phi=\left(
\begin{array}
[c]{c}%
\phi_{1}\\
\phi_{2}%
\end{array}
\right)  $ with mass matrix $m=\frac{1}{2}\left(  m_{1}+m_{2}\right)
\mathsf{1}+$ $\frac{1}{2}\left(  m_{1}-m_{2}\right)  \sigma_{3}$ in a
gravitational background is
\begin{equation}
(g^{\alpha\beta}D_{\alpha}D_{\beta}-m^{2})\Phi=0 \label{KleinGordon}%
\end{equation}
where $D_{\alpha}$ is the covariant derivative. Since the Christoffel symbols
vanish for $a_{i}$ independent of space time the $D_{\alpha}$ coincide with
$\partial_{\alpha}$. Hence
\begin{equation}
\left(  g^{00}\partial_{0}^{2}+2g^{01}\partial_{0}\partial_{1}+g^{11}%
\partial_{1}^{2}\right)  \Phi-m^{2}\Phi=0. \label{KG2}%
\end{equation}
It is useful at this stage to rewrite the state $\left|  i\right\rangle $ in
terms of the mass eigenstates. 
To be specific, from now on we shall restrict ourselves 
to the neutral Kaon system 
$K_0 -{\overline K}_0$, which is produced by a $\phi$-meson at rest, 
i.e. $K_0 -{\overline K}_0$ in their C.M. frame. 
The CP eigenstates (on choosing a suitable
phase convention for the states $\left|  K_{0}\right\rangle $ \ and $\left|
\overline{K_{0}}\right\rangle $ ) are, in standard notation, $\left|  K_{\pm
}\right\rangle $ with
\begin{equation}
\left|  K_{\pm}\right\rangle =\frac{1}{\sqrt{2}}\left(  \left|  K_{0}%
\right\rangle \pm\left|  \overline{K_{0}}\right\rangle \right)  . \label{cp}%
\end{equation}
The mass eigensates $\left|  K_{S}\right\rangle $ and $\left|  K_{L}%
\right\rangle $ are written in terms of $\left|  K_{\pm}\right\rangle $ as%

\begin{equation}
\left|  K_{L}\right\rangle =\frac{1}{\sqrt{1+\left|  \varepsilon_{2}\right|
^{2}}}\left[  \left|  K_{-}\right\rangle \,+\varepsilon_{2}\left|
K_{+}\right\rangle \right]  \label{Klong}%
\end{equation}

and
\begin{equation}
\left|  K_{S}\right\rangle =\frac{1}{\sqrt{1+\left|  \varepsilon_{1}\right|
^{2}}}\left[  \left|  K_{+}\right\rangle \,+\varepsilon_{1}\left|
K_{+}\right\rangle \right]  . \label{Kshort}%
\end{equation}

In terms of the mass eigenstates
\begin{equation}
\left|  i\right\rangle =\mathcal{C}\left\{
\begin{array}
[c]{c}%
\left(  \left|  K_{L}\left(  \overrightarrow{k}\right)  \right\rangle \left|
K_{S}\left(  -\overrightarrow{k}\right)  \right\rangle -\left|  K_{S}\left(
\overrightarrow{k}\right)  \right\rangle \left|  K_{L}\left(  -\overrightarrow
{k}\right)  \right\rangle \right)  +\\
\omega\left(  \left|  K_{S}\left(  \overrightarrow{k}\right)  \right\rangle
\left|  K_{S}\left(  -\overrightarrow{k}\right)  \right\rangle -\left|
K_{L}\left(  \overrightarrow{k}\right)  \right\rangle \left|  K_{L}\left(
-\overrightarrow{k}\right)  \right\rangle \right)
\end{array}
\right\}  \label{CPTV}%
\end{equation}
where $\mathcal{C=}\frac{\sqrt{\left(  1+\left|  \varepsilon_{1}\right|
^{2}\right)  \left(  1+\left|  \varepsilon_{2}\right|  ^{2}\right)  }}%
{\sqrt{2}\left(  1-\varepsilon_{1}\varepsilon_{2}\right)  }$ \cite{Bernabeu}.
In the notation of two level systems (on suppressing the $\overrightarrow{k}$
label) we write%
\begin{align}
\left|  K_{L}\right\rangle  &  =\left|  \uparrow\right\rangle \\
\left|  K_{S}\right\rangle  &  =\left|  \downarrow\right\rangle .\nonumber
\end{align}

\bigskip The unnormalised state \ $\left|  i\right\rangle $ will then be an
\ example of an initial state
\begin{equation}
\left|  \psi\right\rangle =\left|  k,\uparrow\right\rangle ^{\left(  1\right)
}\left|  -k,\downarrow\right\rangle ^{\left(  2\right)  }-\left|
k,\downarrow\right\rangle ^{\left(  1\right)  }\left|  -k,\uparrow
\right\rangle ^{\left(  2\right)  }+\xi\left|  k,\uparrow\right\rangle
^{\left(  1\right)  }\left|  -k,\uparrow\right\rangle ^{\left(  2\right)
}+\xi^{\prime}\left|  k,\downarrow\right\rangle ^{\left(  1\right)  }\left|
-k,\downarrow\right\rangle ^{\left(  2\right)  } \label{initstate}%
\end{equation}
where $\left|  M_{L}\left(  \overrightarrow{k}\right)  \right\rangle =\left|
k,\uparrow\right\rangle $ and we have taken $\overrightarrow{k}$ to have only
a non-zero component $k$ in the $x$-direction; superscripts label the two
separated detectors of the collinear meson pair, $\xi$ and $\xi^{\prime}$ are
complex constants and we have left the state $\left|  \psi\right\rangle $
unnormalised. The evolution of this state is governed by a hamiltonian
$\widehat{H}$
\begin{equation}
\widehat{H}=g^{01}\left(  g^{00}\right)  ^{-1}\widehat{k}-\left(
g^{00}\right)  ^{-1}\sqrt{\left(  g^{01}\right)  ^{2}{k}^{2}-g^{00}\left(
g^{11}k^{2}+m^{2}\right)  } \label{GenKG}%
\end{equation}
which is the natural generalisation of the standard Klein Gordon hamiltonian
in a one particle situation. Moreover $\widehat{k}\left|  \pm k,\uparrow
\right\rangle =\pm k\left|  k,\uparrow\right\rangle $ together with the
corresponding relation for $\downarrow$.

\subsection{Gravitationally-dressed initial entangled state: stationary
perturbation theory and order of magnitude estimates of the $\omega$-effect.}

The effect of space-time foam on the initial entangled state of two neutral
mesons is conceptually difficult to isolate, given that the meson state is
itself entangled with the bath. Nevertheless, in the context of our specific
model, which is written as a stochastic hamiltonian, one can estimate the
order of the associated $\omega$-effect of \cite{Bernabeu} by applying
non-degenerate perturbation theory to the states $\left|  k,\uparrow
\right\rangle ^{\left(  i\right)  }$, $\left|  k,\downarrow\right\rangle
^{\left(  i\right)  } $, $i=1,2$. Although it would be more rigorous to
consider the corresponding density matrices, traced over the unobserved
gravitational degrees of freedom, in order to obtain estimates it will suffice
formally to work with single-meson state vectors) .

Owing to the form of the hamiltonian (\ref{GenKG}) the gravitationally
perturbed states will still be momentum eigenstates. The dominant features of
a possible $\omega$-effect can be seen from a term $\widehat{H_{I}}$ in the
single-particle interaction hamiltonian
\begin{equation}
\widehat{H_{I}} = -\left(  { r_{1} \sigma_{1} + r_{2} \sigma_{2}} \right)
\widehat{k} \label{inthamil}%
\end{equation}
which is the leading order contribution in the small parameters $r_{i}$ (c.f.
(\ref{metric}),(\ref{GenKG})) in $H$ (i.e. $\sqrt{\Delta_{i} }$ are small). 
This implies a modification of the mass eigenstates by the medium of 
quantum gravity, in analogy with the celebrated MSW effect of 
neutrino propagating in matter media~\cite{MSW}, 
but with the important difference here that the effects of the medium 
are directional.

In
first order in perturbation theory the gravitational dressing of $\left|  {k,
\downarrow} \right\rangle ^{\left(  i \right)  } $ leads to a state:%

\begin{equation}
\left|  k^{\left(  i\right)  },\downarrow\right\rangle _{QG}^{\left(
i\right)  } = \left|  k^{\left(  i\right)  },\downarrow\right\rangle ^{\left(
i\right)  } + \left|  k^{\left(  i\right)  },\uparrow\right\rangle ^{\left(
i\right)  } \alpha^{\left(  i\right)  }%
\end{equation}
where
\begin{equation}
\alpha^{\left(  i\right)  }= \frac{^{\left(  i\right)  }\left\langle \uparrow,
k^{\left(  i\right)  }\right|  \widehat{H_{I}}\left|  k^{\left(  i\right)  },
\downarrow\right\rangle ^{\left(  i\right)  }}{E_{2} - E_{1}} \label{qgpert}%
\end{equation}
and correspondingly for $\left|  { k^{\left(  i\right)  }, \uparrow}
\right\rangle ^{\left(  i \right)  } $ the dressed state is obtained from
\ref{qgpert} by $\left|  \downarrow\right\rangle \leftrightarrow\left|
\uparrow\right\rangle $ and $\alpha\to\beta$ where
\begin{equation}
\beta^{\left(  i\right)  }= \frac{^{\left(  i\right)  }\left\langle
\downarrow, k^{\left(  i\right)  }\right|  \widehat{H_{I}}\left|  k^{\left(
i\right)  }, \uparrow\right\rangle ^{\left(  i\right)  }}{E_{1} - E_{2}}
\label{qgpert2}%
\end{equation}
Here the quantities $E_{i} = (m_{i}^{2} + k^{2})^{1/2}$ denote the energy
eigenvalues, and $i=1$ is associated with the up state and $i=2$ with the down
state. With this in mind the totally antisymmetric ``gravitationally-dressed''
state can be expressed in terms of the unperturbed single-particle states as:%

\[
\begin{array}
[c]{l}%
\left|  {k, \uparrow} \right\rangle _{QG}^{\left(  1 \right)  } \left|  { - k,
\downarrow} \right\rangle _{QG}^{\left(  2 \right)  } - \left|  {k,
\downarrow} \right\rangle _{QG}^{\left(  1 \right)  } \left|  { - k, \uparrow}
\right\rangle _{QG}^{\left(  2 \right)  } =\\
\left|  {k, \uparrow} \right\rangle ^{\left(  1 \right)  } \left|  { - k,
\downarrow} \right\rangle ^{\left(  2 \right)  } - \left|  {k, \downarrow}
\right\rangle ^{\left(  1 \right)  } \left|  { - k, \uparrow} \right\rangle
^{\left(  2 \right)  }\\
+ \left|  {k, \downarrow} \right\rangle ^{\left(  1 \right)  } \left|  { - k,
\downarrow} \right\rangle ^{\left(  2 \right)  } \left(  {\beta^{\left(  1
\right)  } - \beta^{\left(  2 \right)  } } \right)  + \left|  {k, \uparrow}
\right\rangle ^{\left(  1 \right)  } \left|  { - k, \uparrow} \right\rangle
^{\left(  2 \right)  } \left(  {\alpha^{\left(  2 \right)  } - \alpha^{\left(
1 \right)  } } \right) \\
+ \beta^{\left(  1 \right)  } \alpha^{\left(  2 \right)  } \left|  {k,
\downarrow} \right\rangle ^{\left(  1 \right)  } \left|  { - k, \uparrow}
\right\rangle ^{\left(  2 \right)  } - \alpha^{\left(  1 \right)  }
\beta^{\left(  2 \right)  } \left|  {k, \uparrow} \right\rangle ^{\left(  1
\right)  } \left|  { - k, \downarrow} \right\rangle ^{\left(  2 \right)  }\\
\label{entangl}%
\end{array}
\]
It should be noted that for $r_{i} \propto\delta_{i1} $ the generated
$\omega $-like effect corresponds to the case $\xi= \xi^{\prime}$ in
(\ref{initstate}) since $\alpha^{\left( i \right) } = - \beta^{\left(
i \right) } $, while the $\omega$-effect of \cite{Bernabeu}
(\ref{CPTV}) corresponds to $r_{i} \propto\delta_{i2}$ (and the
generation of $\xi= -\xi^{\prime}$) since $\alpha^{\left( i \right) }
= \beta^{\left( i \right) } .$ 
In the density matrix these cases can
be distinguished by the off-diagonal terms.

These two cases are physically very different. In 
the case of $\phi$-factories, the former 
corresponds to non-definite strangeness  
in the initial state of the neutral Kaons (seen explicitly
when written in terms of $K_0 - {\overline K}_0$), and hence 
strangeness nonconservation in the initial decay of the 
$\phi$-meson,  
while the latter
conserves this quantum number. We remind the reader 
that in a stochastic quantum-gravity situation, strangeness,
or, in that matter, the appropriate quantum number in the case of other 
neutral mesons, is  not necessarily conserved, and this is reflected in the 
above-described general parametrisation of the interaction
Hamiltonian (\ref{inthamil}) in ``flavour'' space. 

As we shall discuss in the next subsection, the (decoherent) time evolution
of these two cases causes the appearence of terms with 
the opposite effects, as far as the quantrum numbers in question
are concerned. Namely, the strangeness-conserving initial state
leads to the appearance of CPT violating terms with a 
strangeness violating form, while an initially strangeness-violating 
combination generates, under evolution in the foam, a 
strangeness-conserving $\omega$-effect of the form proposed in \cite{Bernabeu}.

We next remark that on averaging the density matrix over the random
variables $r_{i}$, we observe that only terms of order $|\omega|^{2}$
will survive, with the order of $|\omega|^{2}$ being
\begin{equation}
|\omega|^{2} = \mathcal{O}\left(  \frac{1}{(E_{1} - E_{2})} (\langle
\downarrow, k |H_{I} |k, \uparrow\rangle)^{2} \right)  = \mathcal{O}\left(
\frac{\Delta_{2} k^{2}}{(E_{1} - E_{2})^{2}} \right)  \sim\frac{\Delta_{2}
k^{2}}{(m_{1} - m_{2})^{2}} \label{omegaorder}%
\end{equation}
for the physically interesting case in which the momenta are of order of the
rest energies.

Recalling (c.f. (\ref{recoil})) that the variance $\Delta_{2}$ (and also 
$\Delta_{1}$) is of the order
of the square of the momentum transfer (in units of the Planck mass scale
$M_{P}$) during the scattering of the single particle state off a
space-time-foam defect, i.e. $\Delta_{2} = \zeta^{2} k^{2}/M_{P}^{2}$, where
$\zeta$ is at present a phenomenological parameter. It cannot be further
determined due to the lack of a complete theory of quantum gravity, which
would in principle determine the order of the momentum transfer. We arrive at
the following estimate of the order of $\omega$ in this model of foam:
\begin{equation}
|\omega|^{2} \sim\frac{\zeta^{2} k^{4}}{M_{P}^{2} (m_{1} - m_{2})^{2}}
\label{order}%
\end{equation}
Consequently for neutral kaons, with momenta of the order of the rest energies
$|\omega| \sim10^{-4} |\zeta |$, whilst for $B$-mesons we have $|\omega|
\sim10^{-6} |\zeta|$. For $1 > \zeta\ge10^{-2}$ these values for $\omega$ are
not far below the sensitivity of current facilities, such DA$\Phi$NE, and
$\zeta$ may be constrained by future data owing to upgrades of
the DA$\Phi$NE facility or a Super B factory. If the universality of quantum
gravity is assumed then $\zeta$ can also be restricted by data from other
sensitive probes, such as terrestrial and extraterrestrial
neutrinos~\cite{NickSarben}.

\subsection{Time-evolution generated $\omega$-like effects}

We next discuss a similar CPT violating (CPTV) effect generated by the time
evolution of the system. The evolution of the two particle state $\left|
\psi\right\rangle $ is given by
\begin{equation}
\left|  \psi\left(  t\right)  \right\rangle =\exp\left[  -i\left(  \widehat
{H}^{(1)}+\widehat{H}^{\left(  2\right)  }\right)  \frac{t}{\hbar}\right]
\left|  \psi\right\rangle \label{psit}%
\end{equation}
where the superscripts on the $\widehat{H}$ indicate that part of the two
particle state that is being acted on by $\widehat{H}$. On using (\ref{evol1})
and (\ref{evol2}) of Appendix A we find
\begin{equation}
\left|  {\psi\left(  t\right)  }\right\rangle =e^{-i\left(  {\lambda
_{0}^{\left(  1\right)  }+\lambda_{0}^{\left(  2\right)  }}\right)
t}\left\{
\begin{array}
[c]{l}%
\hat{a}_{\uparrow\downarrow}\left(  t\right)  \left|  {k,\uparrow
}\right\rangle ^{\left(  1\right)  }\left|  {-k,\downarrow}\right\rangle
^{\left(  2\right)  }+\hat{a}_{\uparrow\uparrow}\left(  t\right)  \left|
{k,\uparrow}\right\rangle ^{\left(  1\right)  }\left|  {-k,\uparrow
}\right\rangle ^{\left(  2\right)  }\\
+\hat{a}_{\downarrow\downarrow}\left(  t\right)  \left|  {k,\downarrow
}\right\rangle ^{\left(  1\right)  }\left|  {-k,\downarrow}\right\rangle
^{\left(  2\right)  }+\hat{a}_{\downarrow\uparrow}\left(  t\right)  \left|
{k,\downarrow}\right\rangle ^{\left(  1\right)  }\left|  {-k,\uparrow
}\right\rangle ^{\left(  2\right)  }%
\end{array}
\right\}  \label{evol3}%
\end{equation}
where
\begin{equation}
\hat{a}_{\uparrow\downarrow}\left(  t\right)  =\hat{f}^{\left(  1\right)
}\left(  t\right)  \hat{f}^{\left(  2\right)  }\left(  t\right)  ^{\ast
}+\hat{g}^{\left(  1\right)  }\left(  t\right)  ^{\ast}\hat{g}^{\left(
2\right)  }\left(  t\right)  -i\xi_{1}\hat{f}^{\left(  1\right)  }\left(
t\right)  \hat{g}^{\left(  2\right)  }\left(  t\right)  -i\xi_{2}%
\hat{g}^{\left(  1\right)  }\left(  t\right)  ^{\ast}\hat{f}^{\left(
2\right)  }\left(  t\right)  ^{\ast}, \label{evol3a}%
\end{equation}%

\begin{equation}
\hat{a}_{\uparrow\uparrow}\left(  t\right)  =-i\hat{f}^{\left(  1\right)
}\left(  t\right)  \hat{g}^{\left(  2\right)  }\left(  t\right)  ^{\ast
}+i\hat{g}^{\left(  1\right)  }\left(  t\right)  ^{\ast}\hat{f}^{\left(
2\right)  }\left(  t\right)  +\xi_{1}\hat{f}^{\left(  1\right)  }\left(
t\right)  \hat{f}^{\left(  2\right)  }\left(  t\right)  -\xi_{2}%
\hat{g}^{\left(  1\right)  }\left(  t\right)  ^{\ast}\hat{g}^{\left(
2\right)  }\left(  t\right)  ^{\ast}, \label{evol3b}
\end{equation}

\begin{equation}
\hat{a}_{\downarrow\downarrow}\left(  t\right)  =-i\hat{g}^{\left(  1\right)
}\left(  t\right)  \hat{f}^{\left(  2\right)  }\left(  t\right)  ^{\ast}%
+i\hat{f}^{\left(  1\right)  }\left(  t\right)  ^{\ast}\hat{g}^{\left(
2\right)  }\left(  t\right)  -\xi_{1}\hat{g}^{\left(  1\right)  }\left(
t\right)  \hat{g}^{\left(  2\right)  }\left(  t\right)  +\xi_{2}\hat
{f}^{\left(  1\right)  }\left(  t\right)  ^{\ast}\hat{f}^{\left(  2\right)
}\left(  t\right)  ^{\ast}, \label{evol3c}
\end{equation}

and%

\begin{equation}
\hat{a}_{\downarrow\uparrow}\left(  t\right)  =-\hat{g}^{\left(  1\right)
}\left(  t\right)  \hat{g}^{\left(  2\right)  }\left(  t\right)  ^{\ast}%
-\hat{f}^{\left(  1\right)  }\left(  t\right)  ^{\ast}\hat{f}^{\left(
2\right)  }\left(  t\right)  -i\xi_{1}\hat{g}^{\left(  1\right)  }\left(
t\right)  \hat{f}^{\left(  2\right)  }\left(  t\right)  -i\xi_{2}\hat
{f}^{\left(  1\right)  }\left(  t\right)  ^{\ast}\hat{g}^{\left(  2\right)
}\left(  t\right)  ^{\ast}. \label{evol3d}
\end{equation}

The operators $\widehat{f}^{\left(  i\right)  },\widehat{f}^{\left(  i\right)
\ast},\widehat{g}^{\left(  i\right)  }$ and $\widehat{g}^{\left(  i\right)
\ast}$ here act on eigenstates and so they produce c-number eigenvalues
$f^{\left(  i\right)  },\,f^{\left(  i\right)  \ast},g^{\left(  i\right)  }$
and $g^{\left(  i\right)  \ast}$ respectively. A similar convention will be
used for other related operators. In the evolution given by (\ref{evol3}) we
shall examine terms involving $\left|  {k,\uparrow}\right\rangle ^{\left(
1\right)  }\left|  {-k,\uparrow}\right\rangle ^{\left(  2\right)  }$ and
$\left|  {k,\downarrow}\right\rangle ^{\left(  1\right)  }\left|
{-k,\downarrow}\right\rangle ^{\left(  2\right)  }$ but independent of
$\xi_{1}$ and $\xi_{2}$. Consequently these may be regarded as a generation of
the $\omega$ contribution as a consequence of entanglement with the
gravitational space-time foam rather than as a modification of an existing
term. Since the hamiltonian generating the evolution is stochastic, the state
of the system at any time is given by $\left(  \left|  \psi\left(  t\right)
\right\rangle \left\langle \psi\left(  t\right)  \right|  \right)  _{avg}$
where the averaging is over the stochastic parameters in the metric. The
resulting state is a mixed density matrix.

Just as in the estimates of the order of the $\omega$-effect for the stationary
states, it is adequate to work in terms of the wavefunctions. The magnitude
of the $\omega$ generated in (\ref{evol3}) will now be estimated. We define
$\varpi$ (\ref{evol3b}) and $\varpi^{\prime}$ (\ref{evol3c}) as
\begin{equation}
\varpi\left(  t\right)  =i\left(  f^{\left(  2\right)  }\left(  t\right)
g^{\left(  1\right)  \ast}\left(  t\right)  -f^{\left(  1\right)  }\left(
t\right)  g^{\left(  2\right)  \ast}\left(  t\right)  \right)
\end{equation}
and
\begin{equation}
\varpi^{\prime}\left(  t\right)  =i\left(  f^{\left(  1\right)  \ast}\left(
t\right)  g^{\left(  2\right)  }\left(  t\right)  -g^{\left(  1\right)
}\left(  t\right)  f^{\left(  2\right)  \ast}\left(  t\right)  \right)  .
\end{equation}
{}From (\ref{fdef}) and (\ref{gdef}) of Appendix A we deduce that
\begin{equation}
\varpi\left(  t\right)  =i\left[  \left(  \cos\left(  \left|  \lambda^{\left(
2\right)  }\right|  t\right)  -i\sin\left(  \left|  \lambda^{\left(  2\right)
}\right|  t\right)  \,\frac{\lambda_{3}^{\left(  2\right)  }}{\left|
\lambda^{\left(  2\right)  }\right|  }\right)  \sin\left(  \left|
\lambda^{\left(  1\right)  }\right|  t\right)  \frac{\lambda_{1}^{\left(
1\right)  }-i\lambda_{2}^{\left(  1\right)  }}{\left|  \lambda^{\left(
1\right)  }\right|  }-\left(  1\leftrightarrow2\right)  \right]
\label{genomega}%
\end{equation}
and
\begin{equation}
\varpi^{\prime}\left(  t\right)  =-i\left[  \left(  \cos\left(  \left|
\lambda^{\left(  2\right)  }\right|  t\right)  +i\sin\left(  \left|
\lambda^{\left(  2\right)  }\right|  t\right)  \,\frac{\lambda_{3}^{\left(
2\right)  }}{\left|  \lambda^{\left(  2\right)  }\right|  }\right)
\sin\left(  \left|  \lambda^{\left(  1\right)  }\right|  t\right)
\frac{\lambda_{1}^{\left(  1\right)  }+i\lambda_{2}^{\left(  1\right)  }%
}{\left|  \lambda^{\left(  1\right)  }\right|  }-\left(  1\leftrightarrow
2\right)  \right]  . \label{genprimeomeg}%
\end{equation}
Clearly we have an exact relation
\begin{equation}
\varpi^{\prime}\left(  t\right)  =\varpi\left(  t\right)  ^{\ast}.
\label{omegreln}%
\end{equation}
Moreover
\begin{align}
\operatorname{Re}\left(  \varpi\left(  t\right)  \right)   &  =\sin\left(
\left|  \lambda^{\left(  2\right)  }\right|  t\right)  \sin\left(  \left|
\lambda^{\left(  1\right)  }\right|  t\right)  \frac{\lambda_{1}^{\left(
1\right)  }\lambda_{3}^{\left(  2\right)  }-\lambda_{1}^{\left(  2\right)
}\lambda_{3}^{\left(  1\right)  }}{\left|  \lambda^{\left(  1\right)
}\right|  \left|  \lambda^{\left(  2\right)  }\right|  }\nonumber\\
&  +\cos\left(  \left|  \lambda^{\left(  2\right)  }\right|  t\right)
\sin\left(  \left|  \lambda^{\left(  1\right)  }\right|  t\right)
\frac{\lambda_{2}^{\left(  1\right)  }}{\left|  \lambda^{\left(  1\right)
}\right|  }-\cos\left(  \left|  \lambda^{\left(  1\right)  }\right|  t\right)
\sin\left(  \left|  \lambda^{\left(  2\right)  }\right|  t\right)
\frac{\lambda_{2}^{\left(  2\right)  }}{\left|  \lambda^{\left(  2\right)
}\right|  } \label{Reomega}%
\end{align}
and%
\begin{align}
\operatorname{Im}\left(  \varpi\left(  t\right)  \right)   &  =\sin\left(
\left|  \lambda^{\left(  2\right)  }\right|  t\right)  \sin\left(  \left|
\lambda^{\left(  1\right)  }\right|  t\right)  \frac{\lambda_{2}^{\left(
2\right)  }\lambda_{3}^{\left(  1\right)  }-\lambda_{2}^{\left(  1\right)
}\lambda_{3}^{\left(  2\right)  }}{\left|  \lambda^{\left(  1\right)
}\right|  \left|  \lambda^{\left(  2\right)  }\right|  }\nonumber\\
&  +\cos\left(  \left|  \lambda^{\left(  2\right)  }\right|  t\right)
\sin\left(  \left|  \lambda^{\left(  1\right)  }\right|  t\right)
\frac{\lambda_{1}^{\left(  1\right)  }}{\left|  \lambda^{\left(  1\right)
}\right|  }-\cos\left(  \left|  \lambda^{\left(  1\right)  }\right|  t\right)
\sin\left(  \left|  \lambda^{\left(  2\right)  }\right|  t\right)
\frac{\lambda_{1}^{\left(  2\right)  }}{\left|  \lambda^{\left(  2\right)
}\right|  }. \label{Imomega}%
\end{align}
To lowest order ( in the strength of the metric fluctuations)
\begin{equation}
\lambda_{3}^{\left(  2\right)  }\lambda_{1}^{\left(  1\right)  }-\lambda
_{3}^{\left(  1\right)  }\lambda_{1}^{\left(  2\right)  }\approx-2r_{1}%
k\chi_{3} \label{est1}%
\end{equation}
and $\chi_{3}\approx\frac{1}{2}\left(  \sqrt{k^{2}+m_{1}^{2}}-\sqrt
{k^{2}+m_{2}^{2}}\right)  $ (which is small for $\left(  m_{1}-m_{2}\right)
\ll k$) and
\begin{align}
&  \cos\left(  \left|  \lambda^{\left(  2\right)  }\right|  t\right)
\sin\left(  \left|  \lambda^{\left(  1\right)  }\right|  t\right)
\frac{\lambda_{2}^{\left(  1\right)  }}{\left|  \lambda^{\left(  1\right)
}\right|  }-\cos\left(  \left|  \lambda^{\left(  1\right)  }\right|  t\right)
\sin\left(  \left|  \lambda^{\left(  2\right)  }\right|  t\right)
\frac{\lambda_{2}^{\left(  2\right)  }}{\left|  \lambda^{\left(  2\right)
}\right|  }\nonumber\\
&  \approx-\sin\left(  2\left(  \left|  \lambda^{\left(  1\right)  }\right|
+r_{3}k\right)  t\right)  \frac{r_{2}k}{\chi_{3}}\,. \label{est2}%
\end{align}
Hence for $r_{2}\neq0$ $\operatorname{Re}\left(  \varpi\left(  t\right)
\right)  \gg\operatorname{Im}\left(  \varpi\left(  t\right)  \right)  $ and
$\varpi\left(  t\right)  \approx\varpi^{\prime}\left(  t\right)  $. By
contrast when $r_{2}=0$ and $r_{1}\neq0$ $\operatorname{Im}\left(
\varpi\left(  t\right)  \right)  \gg\operatorname{Re}\left(  \varpi\left(
t\right)  \right)  $ and so $\varpi\left(  t\right)  ^{\ast}\sim-\varpi\left(
t\right)  $ which is the permutation symmetry necessary for the $\omega
$-effect~\cite{Bernabeu}. This leads to the simplification
\begin{equation}
\varpi\left(  t\right)  \sim i\cos\left(  \left|  \lambda^{\left(  2\right)
}\right|  t\right)  \sin\left(  \left|  \lambda^{\left(  1\right)  }\right|
t\right)  \left[  \frac{\lambda_{1}^{\left(  1\right)  }}{\left|
\lambda^{\left(  1\right)  }\right|  }-\frac{\lambda_{1}^{\left(  2\right)  }%
}{\left|  \lambda^{\left(  2\right)  }\right|  }\right]  \label{omegapprox}%
\end{equation}
and so the leading contribution to $\left|  \psi\left(  t\right)
\right\rangle $ of the CPT\ violating type is
\begin{equation}
\left|  \psi\left(  t\right)  \right\rangle \sim e^{-i\left(  \lambda
_{0}^{\left(  1\right)  }+\lambda_{0}^{\left(  2\right)  }\right)  t}%
\varpi\left(  t\right)  \left\{  \left|  k,\uparrow\right\rangle ^{\left(
1\right)  }\left|  -k,\uparrow\right\rangle ^{\left(  2\right)  }-\left|
k,\downarrow\right\rangle ^{\left(  1\right)  }\left|  -k,\downarrow
\right\rangle ^{\left(  2\right)  }\right\}  \label{CPTV2}%
\end{equation}
{}From (\ref{omegapprox}) $O\left(  \varpi\right)  \simeq\frac{\lambda
_{1}^{\left(  1\right)  }-\lambda_{1}^{\left(  2\right)  }}{\left|
\lambda^{\left(  1\right)  }\right|  }\sim2\left(  1+\Delta_{4}^{\frac{1}{2}%
}\right)  \frac{\Delta_{1}^{\frac{1}{2}}k}{\left|  \lambda^{\left(  1\right)
}\right|  }$ and $\left|  \lambda^{\left(  1\right)  }\right|  \sim\left(
1+\Delta_{4}^{\frac{1}{2}}\right)  \sqrt{\chi_{1}^{2}+\chi_{3}^{2}}$. From
(\ref{chi01}), (\ref{chi02}) and (\ref{chi03}) of Appendix A we observe that,
to leading order, $\chi_{3}\sim\left(
k^{2}+m_{1}^{2}\right)  ^{\frac{1}{2}}-\left(  k^{2}+m_{2}^{2}\right)
^{\frac{1}{2}}$ and so
\begin{equation}
O\left(  \varpi\right)  \simeq\frac{2\Delta_{1}^{\frac{1}{2}}k}{\left(
k^{2}+m_{1}^{2}\right)  ^{\frac{1}{2}}-\left(  k^{2}+m_{2}^{2}\right)
^{\frac{1}{2}}}\cos\left(  \left|  \lambda^{\left(  1\right)  }\right|
t\right)  \sin\left(  \left|  \lambda^{\left(  1\right)  }\right|  t\right)
=\varpi_{0}\sin\left(  2\left|  \lambda^{\left(  1\right)  }\right|  t\right)
. \label{CPTV3}%
\end{equation}
\bigskip\textbf{ }with $\varpi_{0}\equiv\frac{\Delta_{1}^{\frac{1}{2}}%
k}{\left(  k^{2}+m_{1}^{2}\right)  ^{\frac{1}{2}}-\left(  k^{2}+m_{2}%
^{2}\right)  ^{\frac{1}{2}}}$.

Although $\Delta_{1}$ is a parameter, our model has its origins in models of
D-particle foam; as previously discussed the estimate for $\Delta_{1}$ that
arises in such models is given in terms of the momentum transfer during the
scattering of the matter state with the space-time defect
\[
\Delta_{1}^{1/2}\sim\left|  \zeta\right|  \frac{\left|  k\right|  }{M_{P}}.
\]
This yields an $|\varpi_{0}|$ of the same order as $|\omega|$ in (\ref{order}).

The time dependence of this effect allows its experimental disentanglement
from the $\omega$-effect appearing in the initial state of the two neutral
mesons. The situation should be compared with the analogous one within the
context of a Lindblad approach to the foam, considered in \cite{waldron},
where again the evolution effects can be disentangled from the initial-state
symmetry CPTV $\omega$-effects.

\bigskip\textbf{ }

\section{The Thermal Master Equation}

\bigskip

Master equations with a thermal bath have been argued to be relevant to
decoherence with space time foam \cite{Garay}. A thermal field represents a
bath about which there is minimal information since only the mean energy of
the bath is known, a situation valid also for space time foam. In applications
of quantum information it has been shown that a system of two qubits (or
two-level systems) initially in a separable state (and interacting with a
thermal bath) can actually be entangled by such a single mode
bath~\cite{Entanglement}. As the system evolves the degree of entanglement is
sensitive to the initial state. The close analogy between two-level systems
and neutral meson systems, together with the modelling by a phenomenological
thermal bath of space time foam, makes the study of thermal master equations a
rather intriguing one for the generation of $\omega$. The hamiltonian
$\mathcal{H}$ representing the interaction of two such two-level systems with
a single mode thermal field is
\begin{equation}
\mathcal{H}=\hbar\nu a^{\dag}a+\frac{1}{2}\hbar\Omega\sigma_{3}^{\left(
1\right)  }+\frac{1}{2}\hbar\Omega\sigma_{3}^{\left(  2\right)  }+\hbar
\gamma\sum_{i=1}^{2}\left(  a\sigma_{+}^{\left(  i\right)  }+a^{\dag}%
\sigma_{-}^{\left(  i\right)  }\right)  \label{thermal}%
\end{equation}
where $a$ is the annihilation operator for the mode of the thermal field and
the $\sigma$'s are the standard Pauli matrices for the 2 level systems.The
operators $a$ and $a^{\dag}$ satisfy
\begin{equation}
\left[  a,a^{\dag}\right]  =1,\left[  a^{\dag},a^{\dag}\right]  =\left[
a,a\right]  =0.
\end{equation}
The superscripts label the particle. This hamiltonian is the Jaynes-Cummings
hamiltonian \cite{Jaynes} and explicitly incorporates the back reaction or
entanglement between system and bath. This is in contrast with the Lindlblad
model and the Liouville stochastic metric model. In common with the Lindblad
model it is non-geometric. In the former the entanglement with the bath has
been in principle integrated over while in the latter it is represented by a
stochastic effect. It is useful to go to the interaction picture in which
operators are labelled by a subscript $\mathrm{I;}$ in terms of the
Schr\"{o}dinger picture the interaction picture operators are given by%
\begin{align}
a_{\mathrm{I}}\left(  t\right)   &  =\exp\left(  -i\upsilon t\right)
\,a\nonumber\\
\sigma_{3\mathrm{I}}\left(  t\right)   &  =\sigma_{3}\label{interaction}\\
\sigma_{+\mathrm{I}}\left(  t\right)   &  =\exp\left(  i\Omega t\right)
\sigma_{+}.\nonumber
\end{align}
Moreover the interaction part of the hamiltonian $V=\hbar\gamma\sum_{i=1}%
^{2}\left(  a\sigma_{+}^{\left(  i\right)  }+a^{\dag}\sigma_{-}^{\left(
i\right)  }\right)  $ \thinspace\ transforms to
\begin{equation}
V_{\mathrm{I}}=\hbar\gamma\sum_{i=1}^{2}\left(  \exp\left(  -i\delta t\right)
a\sigma_{+}^{\left(  i\right)  }+\exp\left(  i\delta t\right)  a^{\dag}%
\sigma_{-}^{\left(  i\right)  }\right)  \label{interactionV}%
\end{equation}
where $\delta=\nu-\Omega$. The initial density matrix $\rho\left(  0\right)  $
is taken as
\begin{equation}
\rho\left(  0\right)  =\rho_{M}\otimes\rho_{F}.
\end{equation}
Here
\begin{equation}
\rho_{M}=\left|  \mathfrak{A}\right\rangle \left\langle \mathfrak{A}\right|  ,
\end{equation}%
\begin{equation}
\left|  \mathfrak{A}\right\rangle =\left|  \uparrow\right\rangle ^{\left(
1\right)  }\left|  \downarrow\right\rangle ^{\left(  2\right)  }-\left|
\downarrow\right\rangle ^{\left(  1\right)  }\left|  \uparrow\right\rangle
^{\left(  2\right)  }+\xi_{1}\left|  \uparrow\right\rangle ^{\left(  1\right)
}\left|  \uparrow\right\rangle ^{\left(  2\right)  }+\xi_{2}\left|
\uparrow\right\rangle ^{\left(  1\right)  }\left|  \uparrow\right\rangle
^{\left(  2\right)  },
\end{equation}
and
\begin{equation}
\rho_{F}=\sum_{n=0}^{\infty}\frac{\overline{n}^{n}}{\left(  1+\overline
{n}\right)  ^{n+1}}\left|  n\right\rangle \left\langle n\right|  .
\end{equation}
Hence
\begin{equation}
\rho\left(  0\right)  =\sum_{n=0}^{\infty}\frac{\overline{n}^{n}}{\left(
1+\overline{n}\right)  ^{n+1}}\left|  \mathfrak{A}\right\rangle \left|
n\right\rangle \left\langle n\right|  \left\langle \mathfrak{A}\right|  .
\end{equation}
{}From Bose-Einstein statistics
\begin{equation}
\overline{n}=\frac{1}{-1+e^{\frac{\hbar\nu}{k_{B}T}}} \label{BE}%
\end{equation}
where $T$ is the temperature of the heat bath.The stationary states of the
total matter-bath system are non-separable as noted in Appendix B.
Consequently the effect of the bath cannot be found by means of an effective
hamiltonian (for the matter system on its own) perturbed from the hamiltonian
in the absence of the bath. The dynamics of the matter system on its own is
described by a non-Markovian master equation. However we will describe the
evolution of the matter directly by considering the hamiltonian evolution of
the combined\ matter-bath system and then by tracing over the bath degrees of
freedom. As described in Appendix B the combined evolution takes place in four
dimensional subspaces $\mathcal{S}_{n}$ . Hence even though the full Hilbert
space is infinite dimensional the dynamical evolution can be obtained as the
direct sum of the evolved vectors in each $\mathcal{S}_{n}$. $\left|
\mathfrak{A}\right\rangle \left|  n\right\rangle $ evolves to $\left|
\Phi_{n}\left(  t\right)  \right\rangle $ where
\begin{align}
\left|  \Phi_{n}\left(  t\right)  \right\rangle  &  =X_{1}^{\left(
n-2,4\right)  }\left(  t\right)  \xi_{2}\left|  \uparrow\right\rangle
^{\left(  1\right)  }\left|  \uparrow\right\rangle ^{\left(  2\right)
}\left|  n-2\right\rangle \nonumber\\
&  +\left\{
\begin{array}
[c]{c}%
\left(  X_{1}^{\left(  n-1,2\right)  }\left(  t\right)  -X_{1}^{\left(
n-1,3\right)  }\left(  t\right)  \right)  \left|  \uparrow\right\rangle
^{\left(  1\right)  }\left|  \uparrow\right\rangle ^{\left(  2\right)  }%
+\xi_{2}X_{2}^{\left(  n-2,4\right)  }\left(  t\right)  \left|  \uparrow
\right\rangle ^{\left(  1\right)  }\left|  \downarrow\right\rangle ^{\left(
2\right)  }\\
+\xi_{2}X_{3}^{\left(  n-2,4\right)  }\left(  t\right)  \left|  \downarrow
\right\rangle ^{\left(  1\right)  }\left|  \uparrow\right\rangle ^{\left(
2\right)  }%
\end{array}
\right\}  \left|  n-1\right\rangle \nonumber\\
&  +\left\{
\begin{array}
[c]{c}%
\xi_{1}X_{1}^{\left(  n,1\right)  }\left(  t\right)  \left|  \uparrow
\right\rangle ^{\left(  1\right)  }\left|  \uparrow\right\rangle ^{\left(
2\right)  }+\left(  X_{2}^{\left(  n-1,2\right)  }\left(  t\right)
-X_{2}^{\left(  n-1,3\right)  }\left(  t\right)  \right)  \left|
\uparrow\right\rangle ^{\left(  1\right)  }\left|  \downarrow\right\rangle
^{\left(  2\right)  }\\
+\left(  X_{3}^{\left(  n-1,2\right)  }\left(  t\right)  -X_{3}^{\left(
n-1,3\right)  }\left(  t\right)  \right)  \left|  \downarrow\right\rangle
^{\left(  1\right)  }\left|  \uparrow\right\rangle ^{\left(  2\right)  }%
+\xi_{2}X_{4}^{\left(  n-2,4\right)  }\left(  t\right)  \left|  \downarrow
\right\rangle ^{\left(  1\right)  }\left|  \downarrow\right\rangle ^{\left(
2\right)  }%
\end{array}
\right\}  \left|  n\right\rangle \nonumber\\
&  +\left\{
\begin{array}
[c]{c}%
\xi_{1}X_{2}^{\left(  n,1\right)  }\left(  t\right)  \left|  \uparrow
\right\rangle ^{\left(  1\right)  }\left|  \downarrow\right\rangle ^{\left(
2\right)  }+\xi_{1}X_{3}^{\left(  n,1\right)  }\left(  t\right)  \left|
\downarrow\right\rangle ^{\left(  1\right)  }\left|  \uparrow\right\rangle
^{\left(  2\right)  }\\
+\left(  X_{4}^{\left(  n-1,2\right)  }\left(  t\right)  -X_{4}^{\left(
n-1,3\right)  }\left(  t\right)  \right)  \left|  \downarrow\right\rangle
^{\left(  1\right)  }\left|  \downarrow\right\rangle ^{\left(  2\right)  }%
\end{array}
\right\}  \left|  n+1\right\rangle \nonumber\\
&  +\xi_{1}X_{4}^{\left(  n,1\right)  }\left(  t\right)  \left|
\downarrow\right\rangle ^{\left(  1\right)  }\left|  \downarrow\right\rangle
^{\left(  2\right)  }\left|  n+2\right\rangle . \label{evol4}%
\end{align}
Consequently at time $t$ the density matrix for the matter by itself is
\begin{equation}
\rho_{M}\left(  t\right)  =\sum_{l=0}^{\infty}\sum_{n=0}^{\infty}%
\frac{\overline{n}^{n}}{\left(  1+\overline{n}\right)  ^{n+1}}\,\left\langle
l\right.  \left|  \Phi_{n}\left(  t\right)  \right\rangle \left\langle
\Phi_{n}\left(  t\right)  \right|  \left.  l\right\rangle .
\label{densitymatrix}%
\end{equation}
We are interested primarily in the terms in $\rho_{M}\left(  t\right)  $ which
involve $\left|  \uparrow\right\rangle ^{\left(  1\right)  }\left|
\uparrow\right\rangle ^{\left(  2\right)  }$ and \ $\left|  \downarrow
\right\rangle ^{\left(  1\right)  }\left|  \downarrow\right\rangle ^{\left(
2\right)  }$. If we denote these terms by $\rho_{\shortparallel}$then it is
straightforward to show that
\begin{align*}
\rho_{\shortparallel}\left(  t\right)   &  =\sum_{n=1}^{\infty}\frac
{\overline{n}^{n}}{\left(  1+\overline{n}\right)  ^{n+1}}\left|
X_{1}^{\left(  n-1,2\right)  }\left(  t\right)  -X_{1}^{\left(  n-1,3\right)
}\left(  t\right)  \right|  ^{2}\left|  \uparrow\right\rangle ^{\left(
1\right)  }\left|  \uparrow\right\rangle ^{\left(  2\right)  }\,^{\left(
2\right)  }\left\langle \uparrow\right|  ^{\left(  1\right)  }\left\langle
\uparrow\right| \\
&  +\sum_{n=0}^{\infty}\frac{\overline{n}^{n}}{\left(  1+\overline{n}\right)
^{n+1}}\left|  X_{4}^{\left(  n-1,2\right)  }\left(  t\right)  -X_{4}^{\left(
n-1,3\right)  }\left(  t\right)  \right|  ^{2}\left|  \downarrow\right\rangle
^{\left(  1\right)  }\left|  \downarrow\right\rangle ^{\left(  2\right)
}\,^{\left(  2\right)  }\left\langle \downarrow\right|  ^{\left(  1\right)
}\left\langle \downarrow\right| \\
&  +\left|  \xi_{2}\right|  ^{2}\sum_{n=2}^{\infty}\frac{\overline{n}^{n}%
}{\left(  1+\overline{n}\right)  ^{n+1}}\left|  X_{1}^{\left(  n-2,4\right)
}\left(  t\right)  \right|  ^{2}\left|  \uparrow\right\rangle ^{\left(
1\right)  }\left|  \uparrow\right\rangle ^{\left(  2\right)  }\,^{\left(
2\right)  }\left\langle \uparrow\right|  ^{\left(  1\right)  }\left\langle
\uparrow\right| \\
&  +\left|  \xi_{1}\right|  ^{2}\sum_{n=0}^{\infty}\frac{\overline{n}^{n}%
}{\left(  1+\overline{n}\right)  ^{n+1}}\left|  X_{4}^{\left(  n,1\right)
}\left(  t\right)  \right|  ^{2}\left|  \downarrow\right\rangle ^{\left(
1\right)  }\left|  \downarrow\right\rangle ^{\left(  2\right)  }\,^{\left(
2\right)  }\left\langle \downarrow\right|  ^{\left(  1\right)  }\left\langle
\downarrow\right| \\
&  +\sum_{n=0}^{\infty}\frac{\overline{n}^{n}}{\left(  1+\overline{n}\right)
^{n+1}}\left(
\begin{array}
[c]{c}%
\left|  \xi_{1}\right|  ^{2}\left|  X_{1}^{\left(  n,1\right)  }\left(
t\right)  \right|  ^{2}\left|  \uparrow\right\rangle ^{\left(  1\right)
}\left|  \uparrow\right\rangle ^{\left(  2\right)  \left(  2\right)
}\left\langle \uparrow\right|  ^{\left(  1\right)  }\left\langle
\uparrow\right|  +\left|  \xi_{2}\right|  ^{2}\left|  X_{4}^{\left(
n-2,4\right)  }\left(  t\right)  \right|  ^{2}\left(  t\right)  \left|
\downarrow\right\rangle ^{\left(  1\right)  }\left|  \downarrow\right\rangle
^{\left(  2\right)  \left(  2\right)  }\left\langle \downarrow\right|
^{\left(  1\right)  }\left\langle \downarrow\right| \\
+\xi_{1}\xi_{2}^{\ast}\left[  X_{4}^{\left(  n-2,4\right)  }\left(  t\right)
\right]  ^{\ast\;}X_{1}^{\left(  n,1\right)  }\left(  t\right)  \,^{2}\left|
\uparrow\right\rangle ^{\left(  1\right)  }\left|  \uparrow\right\rangle
^{\left(  2\right)  \,\left(  2\right)  }\left\langle \downarrow\right|
^{\left(  1\right)  }\left\langle \downarrow\right| \\
+\xi_{1}^{\ast}\xi_{2}\left[  X_{1}^{\left(  n,1\right)  }\left(  t\right)
\right]  ^{\ast}X_{4}^{\left(  n-2,4\right)  }\left(  t\right)  \left|
\downarrow\right\rangle ^{\left(  1\right)  }\left|  \downarrow\right\rangle
^{\left(  2\right)  \left(  2\right)  }\left\langle \uparrow\right|  ^{\left(
1\right)  }\left\langle \uparrow\right|
\end{array}
\right)  .
\end{align*}
In $\rho_{\shortparallel}\left(  t\right)  $ only the terms $\rho
_{\shortparallel}^{\prime}\left(  t\right)  $
\begin{align}
\rho_{\shortparallel}^{\prime}\left(  t\right)   &  =\sum_{n=1}^{\infty}%
\frac{\overline{n}^{n}}{\left(  1+\overline{n}\right)  ^{n+1}}\left|
X_{1}^{\left(  n-1,2\right)  }\left(  t\right)  -X_{1}^{\left(  n-1,3\right)
}\left(  t\right)  \right|  ^{2}\left|  \uparrow\right\rangle ^{\left(
1\right)  }\left|  \uparrow\right\rangle ^{\left(  2\right)  }\,^{\left(
2\right)  }\left\langle \uparrow\right|  ^{\left(  1\right)  }\left\langle
\uparrow\right| \nonumber\\
&  +\sum_{n=0}^{\infty}\frac{\overline{n}^{n}}{\left(  1+\overline{n}\right)
^{n+1}}\left|  X_{4}^{\left(  n-1,2\right)  }\left(  t\right)  -X_{4}^{\left(
n-1,3\right)  }\left(  t\right)  \right|  ^{2}\left|  \downarrow\right\rangle
^{\left(  1\right)  }\left|  \downarrow\right\rangle ^{\left(  2\right)
}\,^{\left(  2\right)  }\left\langle \downarrow\right|  ^{\left(  1\right)
}\left\langle \downarrow\right|  \label{entanglementgenerated}%
\end{align}
are generated during the evolution from the CPT symmetric part of $\left|
\mathfrak{A}\right\rangle $. In Appendix B we noted that $a_{i}^{\left(
n-1,2\right)  }=a_{i}^{\left(  n-1,3\right)  },\,i=1,\ldots,3$ and so
$X_{1}^{\left(  n-1,2\right)  }\left(  t\right)  =X_{1}^{\left(  n-1,3\right)
}\left(  t\right)  $; by the same reasoning $X_{4}^{\left(  n-1,2\right)
}\left(  t\right)  =X_{4}^{\left(  n-1,3\right)  }\left(  t\right)  $.
Consequently $\rho_{\shortparallel}^{\prime}\left(  t\right)  =0$ and so the
thermal bath model does not generate the entanglement implied by CPTV.

\bigskip

\bigskip

\section{Conclusions}

{} In this work we have discussed two classes of space-time foam models, which,
conceivably, may characterise realistic situations of the (still
elusive) theory of quantum gravity. In one of the models (LSM), inspired by
non-critical string theory models of foam, but placed here in the more general
context of semi-microscopic effective theories of quantum gravity, there is an
appropriate metric distortion caused by the recoil of the space-time defect
(microscopic black hole) during the scattering with the matter probe. The
distortion is such that it connects the different mass eigenstates of neutral
mesons, and is proportional to the momentum transfer of the matter probe
during its scattering with the space-time defect. The latter is assumed to
fluctuate randomly, with a dispersion which at present is viewed as a
phenomenological parameter to be constrained by data.

This causes a CPTV $\omega$-like effect in the initial entangled state of two
neutral mesons in a meson factory, of the type conjectured in \cite{Bernabeu}.
Using stationary (non-degenerate) perturbation theory it was possible to give
an order of magnitude estimate of the effect: the latter is momentum
dependent, and of an order which may not be far from the sensitivity of experimental facilities in the near future, such as a possible upgrade of the DA$\Phi$NE facility or a Super B factory. A similar effect, but with a sinusoidal time
dependence, and hence experimentally distinguishable from the initial-state
effect , is also generated in this model of foam by the evolution of the
system. The situation needs to be compared with Lindblad-type phenomenological
models of quantum gravity foam, where again evolution-generated CPT Violating
effects can be disentangled from the $\omega$-effect characterising the
initial state.

As we have discussed, there are two physically very different
cases for the initial state in a quantum gravity situation of an LSM-like 
model. One involves an initial state of two mesons 
with definte ``strangeness'' (i.e. the appropriate quantum number for 
the meson factory in question), which the space-time foam evolves 
into an indefinite ``strangeness'' combination (i.e. strangeness violation).
The other case corresponds to an indefinite strangeness initial 
state, whose (decoherent) time  evolution yields time-dependent 
$\omega$-effect terms
with definite strangeness of the type 
considered in \cite{Bernabeu}. These two cases are produced by 
the different terms in the initial perturbation Hamiltonian 
(\ref{inthamil}). 

A second model of space-time foam, that of a thermal bath of gravitational
degrees of freedom, is also considered in our work, which, however, does not
lead to the generation of an $\omega$-like effect.

It is interesting to continue the search for more realistic models of quantum
gravity, either in the context of string theory or in other approaches, such
as the canonical approach or the loop quantum gravity, in order to search for
intrinsic CPT violating effects in sensitive matter probes. Detailed analyses
of global data, including very sensitive probes such as high energy neutrinos,
is the only way forward in order to obtain some clues on the elusive theory of
quantum gravity. Decoherence, induced by quantum gravity, far from being
ruled out at present, may indeed provide the link between theory and
experiment in this intriguing area of physics. 

\section*{Acknowledgements}

It is a pleasure to acknowledge informative discussions with
G. Amelino-Camelia, G. Barenboim, 
A. Di Domenico, G. Isidori and J. Papavassiliou. 
The work of J.B. is supported by the Spanish M.E.C. and European 
FEDER under grant FPA 2005-01678. 
N.E.M. wishes to thank the Department of Theoretical Physics 
of the University of Valencia (Spain) for the hospitality 
during the last stages of this work.

\section*{Appendix A: details of LSM evolution}

In order to simplify $\widehat{H}$ (\ref{GenKG}) we note the decomposition
\begin{equation}
\sqrt{\left(  {\widehat{A}\mathsf{1}+\widehat{B}\sigma_{3}+\widehat{C}%
\sigma_{1}+\widehat{D}\sigma_{2}}\right)  }=\hat{\chi_{0}}\mathsf{1}%
+\sum\limits_{j=1}^{3}{\hat{\chi}_{j}\sigma_{j}} \label{decomp}%
\end{equation}
where
\begin{equation}
\hat{\chi}_{0}=\frac{1}{2}\left[  {\left(  {\widehat{A}+\sqrt{\widehat{B}%
^{2}+\widehat{C}^{2}+\widehat{D}^{2}}}\right)  ^{{1\mathord{\left/
{\vphantom{1 2}} \right.  \kern-\nulldelimiterspace}2}}+\left(  {\widehat
{A}-\sqrt{\widehat{B}^{2}+\widehat{C}^{2}+\widehat{D}^{2}}}\right)
^{{1\mathord{\left/  {\vphantom{1 2}} \right.  \kern-\nulldelimiterspace}2}}%
}\right]  , \label{chi00}%
\end{equation}%
\begin{equation}
\hat{\chi}_{1}=\frac{{\widehat{C}}}{{2\sqrt{\widehat{B}^{2}+\widehat{C}%
^{2}+\widehat{D}^{2}}}}\left[  {\left(  {\widehat{A}+\sqrt{\widehat{B}%
^{2}+\widehat{C}^{2}+\widehat{D}^{2}}}\right)  ^{{1\mathord{\left/
{\vphantom{1 2}} \right.  \kern-\nulldelimiterspace}2}}-\left(  {\widehat
{A}-\sqrt{\widehat{B}^{2}+\widehat{C}^{2}+\widehat{D}^{2}}}\right)
^{{1\mathord{\left/  {\vphantom{1 2}} \right.  \kern-\nulldelimiterspace}2}}%
}\right]  , \label{chi01}%
\end{equation}%
\begin{equation}
\hat{\chi}_{2}=\frac{{\hat{D}}}{{2\sqrt{\widehat{B}^{2}+\widehat{C}%
^{2}+\widehat{D}^{2}}}}\left[  {\left(  {\widehat{A}+\sqrt{\widehat{B}%
^{2}+\widehat{C}^{2}+\widehat{D}^{2}}}\right)  ^{{1\mathord{\left/
{\vphantom{1 2}} \right.  \kern-\nulldelimiterspace}2}}-\left(  {\widehat
{A}-\sqrt{\widehat{B}^{2}+\widehat{C}^{2}+\widehat{D}^{2}}}\right)
^{{1\mathord{\left/  {\vphantom{1 2}} \right.  \kern-\nulldelimiterspace}2}}%
}\right]  , \label{chi02}%
\end{equation}
and
\begin{equation}
\hat{\chi}_{3}=\frac{{\widehat{B}}}{{2\sqrt{\widehat{B}^{2}+\widehat{C}%
^{2}+\widehat{D}^{2}}}}\left[  {\left(  {\widehat{A}+\sqrt{\widehat{B}%
^{2}+\widehat{C}^{2}+\widehat{D}^{2}}}\right)  ^{{1\mathord{\left/
{\vphantom{1 2}} \right.  \kern-\nulldelimiterspace}2}}-\left(  {\widehat
{A}-\sqrt{\widehat{B}^{2}+\widehat{C}^{2}+\widehat{D}^{2}}}\right)
^{{1\mathord{\left/  {\vphantom{1 2}} \right.  \kern-\nulldelimiterspace}2}}%
}\right]  . \label{chi03}%
\end{equation}
Using (\ref{decomp}) we can write the single particle hamiltonian as%
\begin{equation}
\widehat{H}=\left(  1+r_{4}\right)  \left\{  -\left(  r_{0}{\mathsf{1}}%
+r_{1}\sigma_{1}+r_{2}\sigma_{2}+r_{3}\sigma_{3}\right)  \widehat{k}+\left(
\hat{\chi}_{0}{\mathsf{1}}+\sum_{j=1}^{3}\hat{\chi}_{j}\sigma_{j}\right)
\right\}
\end{equation}
for
\begin{align}
{\widehat{A}}  &  {=}\widehat{k}^{2}\,\left[  \left(  r_{0}^{2}+r_{1}%
^{2}+r_{2}^{2}+r_{3}^{2}\right)  +\left(  1-r_{4}\right)  \left(
1+r_{2}\right)  \right]  +\frac{1}{2}\left(  1-r_{4}\right)  \left(  m_{1}%
^{2}+m_{2}^{2}\right) \\
\widehat{{B}}  &  {=}{2}\widehat{k}^{2}r_{0}r_{3}+\frac{1}{2}\left(  m_{1}%
^{2}-m_{2}^{2}\right)  \left(  1-r_{4}\right) \\
\widehat{{C}}  &  {=}{2}\widehat{k}^{2}r_{0}r_{1}.
\end{align}
and
\begin{equation}
\widehat{D}={2}\widehat{k}^{2}r_{0}r_{2}.
\end{equation}
The operators ${\widehat{A},}\widehat{B},\widehat{C}$ and $\widehat{D}$ will
always act on eigenstates $\left|  p\right\rangle $ \ (where $\widehat{k}$
$\left|  p\right\rangle =p\left|  p\right\rangle $). The evolution operator
$e^{-i\widehat{H}t}$ in turn has the decomposition
\begin{equation}
e^{-i\widehat{H}t}=e^{-i\widehat{\lambda}_{0}t}e^{-i\sum_{j=1}^{3}%
\widehat{\lambda}_{j}\sigma_{j}t}%
\end{equation}
where
\begin{equation}
\widehat{\lambda}_{\mu}=\left(  1+r_{4}\right)  \left(\hat{\chi}_{\mu}
-r_{\mu}\widehat{k}\right),\,\mu=0,\ldots,3. \label{lambda}%
\end{equation}

Now
\begin{equation}
e^{-i\sum_{j=1}^{3}\widehat{\lambda}_{j}\sigma_{j}t}=\cos\left(  \left|
\widehat{\lambda}\right|  t\right)  -i\sum_{j=1}^{3}\widehat{\lambda}%
_{j}\sigma_{j}\left(  \left|  \widehat{\lambda}\right|  \right)  ^{-1}%
\sin\left(  \left|  \widehat{\lambda}\right|  t\right)  \label{Pauli}%
\end{equation}
where $\left|  \widehat{\lambda}\right|  =\sqrt{\sum_{i=1}^{3}\widehat
{\lambda}_{i}^{2}}$ and again it is assumed that the operator $e^{-i\sum
_{j=1}^{3}\widehat{\lambda}_{j}\sigma_{j}t}$ acts on an eigenstate $\left|
p\right\rangle $ of $\widehat{k}$. Consequently
\begin{equation}
\left|  \widehat{\lambda}\right|  \left|  p\right\rangle =\sqrt{\left(
\lambda_{1}^{2}\left(  p\right)  +\lambda_{2}^{2}\left(  p\right)
+\lambda_{3}^{2}\left(  p\right)  \right)  }\left|  p\right\rangle
\label{modop}%
\end{equation}
with
\begin{equation}
\widehat{\lambda}_{j}\left|  p\right\rangle =\lambda_{j}\left(  p\right)
\left|  p\right\rangle
\end{equation}
for $j=1,\ldots,3.$ Now
\begin{equation}
\sum_{j=1}^{3}\widehat{\lambda}_{j}\sigma_{j}\left|  k,\uparrow\right\rangle
=\left(  \widehat{\lambda}_{1}+i\widehat{\lambda}_{2}\right)  \left|
k,\downarrow\right\rangle +\widehat{\lambda}_{3}\left|  k,\uparrow
\right\rangle ,
\end{equation}%
\begin{equation}
\sum_{j=1}^{3}\widehat{\lambda}_{j}\sigma_{j}\left|  k,\downarrow\right\rangle
=\left(  \widehat{\lambda}_{1}-i\widehat{\lambda}_{2}\right)  \left|
k,\uparrow\right\rangle -\widehat{\lambda}_{3}\left|  k,\downarrow
\right\rangle ,
\end{equation}%
\begin{equation}
\widehat{\lambda}_{j}^{\left(  1\right)  }\left|  k,\alpha\right\rangle
^{\left(  1\right)  }=\left(  1+r_{4}\right)  \left(  \widehat{\chi}_{j}%
-r_{j}k\right)  \left|  k,\alpha\right\rangle ^{\left(  1\right)  },
\end{equation}
and
\begin{equation}
\widehat{\lambda}_{j}^{\left(  2\right)  }\left|  -k,\alpha\right\rangle
^{\left(  2\right)  }=\left(  1+r_{4}\right)  \left(  \widehat{\chi}_{j}%
+r_{j}k\right)  \left|  -k,\alpha\right\rangle ^{\left(  2\right)  }%
\end{equation}
where $\alpha=\uparrow,\downarrow$.

Consequently
\begin{equation}
e^{-i\widehat{H}^{\left(  1\right)  }t}\left|  k,\uparrow\right\rangle
^{\left(  1\right)  }=e^{-i\widehat{\lambda}_{0}^{\left(  1\right)  }t}\left(
\widehat{f}^{\left(  1\right)  }\left(  t\right)  \left|  k,\uparrow
\right\rangle ^{\left(  1\right)  }-i\widehat{g}^{\left(  1\right)  }\left(
t\right)  \left|  k,\downarrow\right\rangle ^{\left(  1\right)  }\right)
\label{evol1}%
\end{equation}
and
\begin{equation}
e^{-i\widehat{H}^{\left(  1\right)  }t}\left|  k,\downarrow\right\rangle
^{\left(  1\right)  }=e^{-i\widehat{\lambda}_{0}^{\left(  1\right)  }t}\left(
\widehat{f}^{\left(  1\right)  \ast}\left(  t\right)  \left|  k,\uparrow
\right\rangle ^{\left(  1\right)  }-i\widehat{g}^{\left(  1\right)  }\left(
t\right)  \left|  k,\downarrow\right\rangle ^{\left(  1\right)  }\right)
\label{evol2}%
\end{equation}
where for $j=1,2$
\begin{equation}
\widehat{f}^{\left(  j\right)  }\left(  t\right)  =\cos\left(  \left|
\widehat{\lambda}^{\left(  j\right)  }\right|  t\right)  -i\sin\left(  \left|
\widehat{\lambda}^{\left(  j\right)  }\right|  t\right)  \left|
\widehat{\lambda}^{\left(  j\right)  }\right|  ^{-1}\widehat{\lambda}%
_{3}^{\left(  j\right)  } \label{fdef}%
\end{equation}
and
\begin{equation}
\widehat{g}^{\left(  j\right)  }\left(  t\right)  =\left(  \widehat{\lambda
}_{1}^{\left(  j\right)  }+i\widehat{\lambda}_{2}^{\left(  j\right)  }\right)
\left|  \widehat{\lambda}^{\left(  j\right)  }\right|  ^{-1}\sin\left(
\left|  \widehat{\lambda}^{\left(  j\right)  }\right|  t\right)  .
\label{gdef}%
\end{equation}

$\bigskip$

$\widehat{f}^{\left(  j\right)  \ast}\left(  t\right)  \;$ $\left(
\widehat{g}^{\left(  j\right)  \ast}\left(  t\right)  \right)  $ and
$\widehat{f}^{\left(  j\right)  }\left(  t\right)  $ $\left(  \widehat
{g}^{\left(  j\right)  }\left(  t\right)  \right)  $ share the same
eigenvectors but with complex conjugate eigenvalues. The formulae (\ref{evol1})
and (\ref{evol2}) hold for states also with superfix $2$ when $k\rightarrow-k$.

\bigskip

\section*{Appendix B: details of thermal bath evolution}

\bigskip

In the analysis of the eigenstates of the Jaynes-Cummings hamiltonian
$\mathcal{H}$ it is clear that certain subspaces $\mathcal{S}_{n}$ of the
Hilbert space $\emph{H}$ for the states of the system and bath are invariant
under the action of $\mathcal{H}$. Moreover $\cup_{n}\mathcal{S}_{n}=\emph{H}%
$. $\mathcal{S}_{n}$ is characterised by states of the form
\begin{equation}
\left|  \Psi\left(  t\right)  \right\rangle =X_{1}^{n}\left(  t\right)
\left|  \uparrow,\uparrow,n\right\rangle +X_{2}^{n}\left(  t\right)  \left|
\uparrow,\downarrow,n+1\right\rangle +X_{3}^{n}\left(  t\right)  \left|
\downarrow,\uparrow,n+1\right\rangle +X_{4}^{n}\left(  t\right)  \left|
\downarrow,\downarrow,n+2\right\rangle . \label{subspace}%
\end{equation}
We shall work in the interaction picture but eschew the cumbersome subscript
$\mathrm{I}$. The evolution of the $X_{i}^{n}$ is governed by
\begin{align}
i\,\dot{X}_{1}^{n}\left(  t\right)   &  =\gamma\sqrt{n+1}\exp\left(  -i\delta
t\right)  \left(  X_{2}^{n}\left(  t\right)  +X_{3}^{n}\left(  t\right)
\right)  ,\nonumber\\
i\,\dot{X}_{2}^{n}\left(  t\right)   &  =\gamma\left(  \exp\left(  -i\delta
t\right)  \sqrt{n+2}X_{4}^{n}\left(  t\right)  +\exp\left(  i\delta t\right)
\sqrt{n+1}X_{1}^{n}\left(  t\right)  \right)  ,\label{interaction0}\\
i\,\dot{X}_{3}^{n}\left(  t\right)   &  =\gamma\left(  \exp\left(  -i\delta
t\right)  \sqrt{n+2}X_{4}^{n}\left(  t\right)  +\exp\left(  i\delta t\right)
\sqrt{n+1}X_{1}^{n}\left(  t\right)  \right)  ,\nonumber\\
i\,\dot{X}_{4}^{n}\left(  t\right)   &  =\gamma\exp\left(  i\delta t\right)
\sqrt{n+2}\left(  X_{2}^{n}\left(  t\right)  +X_{3}^{n}\left(  t\right)
\right)  .\nonumber
\end{align}
On writing
\begin{align}
y_{5}^{n}\left(  t\right)   &  =X_{2}^{n}\left(  t\right)  +X_{3}^{n}\left(
t\right) \nonumber\\
y_{1}^{n}\left(  t\right)   &  =\sqrt{n+1}X_{1}^{n}\left(  t\right) \\
y_{4}^{n}\left(  t\right)   &  =\sqrt{n+2}X_{4}^{n}\left(  t\right) \nonumber
\end{align}

we have from (\ref{interaction0})
\begin{align}
i\dot{y}_{1}^{n}\left(  t\right)   &  =\gamma\left(  n+1\right)  \exp\left(
-i\delta t\right)  y_{5}^{n}\left(  t\right)  ,\nonumber\\
i\dot{y}_{4}^{n}\left(  t\right)   &  =\gamma\left(  n+2\right)  \exp\left(
i\delta t\right)  y_{5}^{n}\left(  t\right)  ,\label{interaction1}\\
i\dot{y}_{5}^{n}\left(  t\right)   &  =2\gamma\left(  \exp\left(  -i\delta
t\right)  y_{4}^{n}\left(  t\right)  +\exp\left(  i\delta t\right)  y_{1}%
^{n}\left(  t\right)  \right)  .\nonumber
\end{align}
Any solution of these equations satisfies%
\begin{equation}
i\dddot{y}_{5}^{n}\left(  t\right)  +i\left(  \delta^{2}+\left(  4n+6\right)
\gamma^{2}\right)  \dot{y}_{5}^{n}\left(  t\right)  +2\delta\gamma^{2}%
y_{5}^{n}\left(  t\right)  =0. \label{reduction}%
\end{equation}
Hence $y_{5}^{n}$ $\propto e^{i\lambda^{\left(  n\right)  }t}$ where
$\lambda=\lambda^{\left(  n\right)  }$ is a solution of
\begin{equation}
\lambda^{3}-\left(  \delta^{2}+(4n+6)\gamma^{2}\right)  \lambda+2\delta
\gamma^{2}=0.
\end{equation}
If we denote the three solutions by $\lambda_{i}^{\left(  n\right)  }\left(
i=1,\ldots,3\right)  $ then the system of equations of (\ref{interaction0}) has
a solution
\begin{align}
X_{1}^{n}\left(  t\right)   &  =-\gamma\sqrt{n+1}\left(  \frac{\mathfrak{a}%
_{1}^{\left(  n\right)  }e^{i\left(  \lambda_{1}^{\left(  n\right)  }%
-\delta\right)  t}}{\lambda_{1}^{\left(  n\right)  }-\delta}+\frac
{\mathfrak{a}_{2}^{\left(  n\right)  }e^{i\left(  \lambda_{2}^{\left(
n\right)  }-\delta\right)  t}}{\lambda_{2}^{\left(  n\right)  }-\delta}%
+\frac{\mathfrak{a}_{3}^{\left(  n\right)  }e^{i\left(  \lambda_{3}^{\left(
n\right)  }-\delta\right)  t}}{\lambda_{3}^{\left(  n\right)  }-\delta}\right)
\nonumber\\
X_{2}^{n}\left(  t\right)   &  =\frac{1}{2}\mathfrak{b}_{2}+\frac{1}{2}\left(
\mathfrak{a}_{1}^{\left(  n\right)  }e^{i\lambda_{1}^{\left(  n\right)  }%
t}+\mathfrak{a}_{2}^{\left(  n\right)  }e^{i\lambda_{2}^{\left(  n\right)  }%
t}+\mathfrak{a}_{3}^{\left(  n\right)  }e^{i\lambda_{3}^{\left(  n\right)  }%
t}\right) \label{GenSoln}\\
X_{3}^{n}\left(  t\right)   &  =-\frac{1}{2}\mathfrak{b}_{2}+\frac{1}%
{2}\left(  \mathfrak{a}_{1}^{\left(  n\right)  }e^{i\lambda_{1}^{\left(
n\right)  }t}+\mathfrak{a}_{2}^{\left(  n\right)  }e^{i\lambda_{2}^{\left(
n\right)  }t}+\mathfrak{a}_{3}^{\left(  n\right)  }e^{i\lambda_{3}^{\left(
n\right)  }t}\right) \nonumber\\
X_{4}^{n}\left(  t\right)   &  =-\gamma\sqrt{n+2}\left(  \frac{\mathfrak{a}%
_{1}^{\left(  n\right)  }}{\lambda_{1}^{\left(  n\right)  }+\delta}e^{i\left(
\lambda_{1}^{\left(  n\right)  }+\delta\right)  t}+\frac{\mathfrak{a}%
_{2}^{\left(  n\right)  }}{\lambda_{2}^{\left(  n\right)  }+\delta}e^{i\left(
\lambda_{2}^{\left(  n\right)  }+\delta\right)  t}+\frac{\mathfrak{a}%
_{3}^{\left(  n\right)  }}{\lambda_{3}^{\left(  n\right)  }+\delta}e^{i\left(
\lambda_{3}^{\left(  n\right)  }+\delta\right)  t}\right) \nonumber
\end{align}
The constants $\mathfrak{a}_{i}^{\left(  n\right)  }$ and $\mathfrak{b}_{2}$
are determined by initial conditions. A set of independent initial conditions
is given by $X_{j}^{n}\left(  0\right)  =\delta_{jk}\,\left(  k=1,\ldots
,4\right)  $; the set of associated solutions is denoted by $a_{i}^{\left(
n,k\right)  }$, $b_{2}^{\left(  k\right)  }$ and $X_{i}^{\left(  n,k\right)
}\left(  t\right)  $. Also for consistency we define for $k=1,\ldots,4$
\begin{equation}
X_{1}^{\left(  -1,k\right)  }=X_{1}^{\left(  -2,k\right)  }\,=X_{2}^{\left(
-2,k\right)  }=X_{3}^{\left(  -2,k\right)  }=0.
\end{equation}
The $\lambda_{i}^{\left(  n\right)  }$ are related to the energy eigenvalues
of stationary states of the combined system and bath.\ Owing to the
entanglement between the bath and the matter it is not useful to consider the
bath as leading to a small perturbation of the matter state.

{}From (\ref{interaction0}) and (\ref{GenSoln}) we have for $k=1$%
\begin{align*}
1  &  =-\gamma\sqrt{n+1}\left(  \frac{a_{1}^{\left(  n,1\right)  }}%
{\lambda_{1}^{\left(  n\right)  }-\delta}+\frac{a_{2}^{\left(  n,1\right)  }%
}{\lambda_{2}^{\left(  n\right)  }-\delta}+\frac{a_{3}^{\left(  n,1\right)  }%
}{\lambda_{3}^{\left(  n\right)  }-\delta}\right)  ,\\
0  &  =a_{1}^{\left(  n,1\right)  }+a_{2}^{\left(  n,1\right)  }%
+a_{3}^{\left(  n,1\right)  },\\
0  &  =\frac{a_{1}^{\left(  n,1\right)  }}{\lambda_{1}^{\left(  n\right)
}+\delta}+\frac{a_{2}^{\left(  n,1\right)  }}{\lambda_{2}^{\left(  n\right)
}+\delta}+\frac{a_{3}^{\left(  n,1\right)  }}{\lambda_{3}^{\left(  n\right)
}+\delta},\\
b_{2}^{\left(  1\right)  }  &  =0.
\end{align*}
Similarly for $k=2$%
\begin{align*}
0  &  =\frac{a_{1}^{\left(  n,2\right)  }}{\lambda_{1}^{\left(  n\right)
}-\delta}+\frac{a_{2}^{\left(  n,1\right)  }}{\lambda_{2}^{\left(  n\right)
}-\delta}+\frac{a_{3}^{\left(  n,1\right)  }}{\lambda_{3}^{\left(  n\right)
}-\delta},\\
1  &  =a_{1}^{\left(  n,2\right)  }+a_{2}^{\left(  n,2\right)  }%
+a_{3}^{\left(  n,2\right)  },\\
0  &  =\frac{a_{1}^{\left(  n,2\right)  }}{\lambda_{1}^{\left(  n\right)
}+\delta}+\frac{a_{2}^{\left(  n,2\right)  }}{\lambda_{2}^{\left(  n\right)
}+\delta}+\frac{a_{3}^{\left(  n,2\right)  }}{\lambda_{3}^{\left(  n\right)
}+\delta},\\
b_{2}^{\left(  2\right)  }  &  =0.
\end{align*}
For $k=3$
\begin{align*}
0  &  =\frac{a_{1}^{\left(  n,3\right)  }}{\lambda_{1}^{\left(  n\right)
}-\delta}+\frac{a_{2}^{\left(  n,3\right)  }}{\lambda_{2}^{\left(  n\right)
}-\delta}+\frac{a_{3}^{\left(  n,3\right)  }}{\lambda_{3}^{\left(  n\right)
}-\delta},\\
1  &  =a_{1}^{\left(  n,3\right)  }+a_{2}^{\left(  n,3\right)  }%
+a_{3}^{\left(  n,3\right)  },\\
0  &  =\frac{a_{1}^{\left(  n,3\right)  }}{\lambda_{1}^{\left(  n\right)
}+\delta}+\frac{a_{2}^{\left(  n,3\right)  }}{\lambda_{2}^{\left(  n\right)
}+\delta}+\frac{a_{3}^{\left(  n,3\right)  }}{\lambda_{3}^{\left(  n\right)
}+\delta},\\
b_{2}^{\left(  3\right)  }  &  =-1.
\end{align*}

\bigskip For $k=4$
\begin{align*}
0  &  =\frac{a_{1}^{\left(  n,4\right)  }}{\lambda_{1}^{\left(  n\right)
}-\delta}+\frac{a_{2}^{\left(  n,4\right)  }}{\lambda_{2}^{\left(  n\right)
}-\delta}+\frac{a_{3}^{\left(  n,4\right)  }}{\lambda_{3}^{\left(  n\right)
}-\delta},\\
0  &  =a_{1}^{\left(  n,4\right)  }+a_{2}^{\left(  n,4\right)  }%
+a_{3}^{\left(  n,4\right)  },\\
1  &  =-\gamma\sqrt{n+2}\left(  \frac{a_{1}^{\left(  n,4\right)  }}%
{\lambda_{1}^{\left(  n\right)  }+\delta}+\frac{a_{2}^{\left(  n,4\right)  }%
}{\lambda_{2}^{\left(  n\right)  }+\delta}+\frac{a_{3}^{\left(  n,4\right)  }%
}{\lambda_{3}^{\left(  n\right)  }+\delta}\right)  ,\\
b_{2}^{\left(  4\right)  }  &  =0.
\end{align*}

\end{document}